# The external rhythm of an actor in science:
# New indicators for the science of science

Liming Liang*[1], Ronald Rousseau[2, 3]

*[1] Institute for Science, Technology and Society, Henan Normal University, Xinxiang 453007, China

[2] KU Leuven, MSI, Facultair Onderzoekscentrum ECOOM, Naamsestraat 61, 3000 Leuven, Belgium

ORCID: 0000-0002-3252-2538

[3] Faculty of Social Sciences, University of Antwerp (UA), Middelheimlaan 1, 2020 Antwerp, Belgium;

## Abstract

When calculating citation indicators, whether it is the total number of received citations or the average citations per paper, we always face the same problem: namely, that papers published in different years have varying citation potential. Hence, strictly speaking, their citations cannot be compared. In a former study, we created a new indicator called the "internal rhythm" indicator of an actor. This indicator is constructed using ratios of observed citations to expected citations, taken from the actor's publication-citation matrix. The "internal rhythm" indicator makes it possible to compare the citation performances among different publication years, but it is only valid within the actor-based framework. In this study, we define, create, and explore the “external rhythm” of an actor, which is also a sequence of ratios of observed citations to expected citations. The essential difference between internal rhythm and external rhythm lies in the way they are created and hence in the point of view taken to study an actor. The former is created based on its own publication-citation matrix, while the latter is based on two publication-citation matrices. One is the same as the former. The other one is a publication-citation matrix of a collective, which includes the actor under study. The external rhythm of an actor is a citation-based indicator of research that can be used to compare not only the citation performance of an actor with that of the collective the actor is part of, but also to compare two or more actors within the same collective. We further propose a summary average of ratios indicator.

## Introduction

When conducting a bibliometric study of an entity (such as a university, journal, or country), one common challenge is that papers published in different years have varying citation windows (even if they are of the same length), making citation contributions strictly incomparable. Traditional citation metrics, like total citations or average citations per paper, tend to lose accuracy. To address this issue, one of us developed an indicator called the “internal rhythm” of an entity (Liang, 2005). This indicator enables the assessment of an entity’s citation performance over multiple years. The term “rhythm” as used in these rhythm studies does not refer to ups and downs in time series of bibliometric indicators, but is considered an internal attribute of the evolutionary process of science. It is a combination of what one observes and what one expects.

A basic publication–citation matrix, in short: p–c matrix, is a table showing publication and citation data needed for the calculation of an informetric indicator such as a journal impact factor (Moed et al., 1983, 1985; Rousseau et al., 2018). In general, a p-c matrix has dimensions n and m, where n and m do not have to be equal. In each case, we assume that a pool of data is given. Typical examples of such pools are the global databases such as Clarivate's Web of Science, Scopus, Dimensions, OpenAlex, or local ones such as SciELO in Latin America, or CNKI in China, or subsets of these, such as all LIS journals and their publications. Examples of the use of p–c matrices can be found in Ingwersen et al. (2001), Frandsen & Rousseau (2005), and Liang (2005). Data in such matrices are usually presented in chronological order, as shown in Table 1. In our investigation, the p-c matrix contains the essential data, from which all other variables, such as $C_k$, $O_i$, and $E_i$ are calculated (definitions of these symbols will follow in Section 2), forming the ratio sequence $R_i = O_i / E_i$ (Liang, 2005).

The field of informetrics consists of different types of studies: those dealing with structural aspects, those dealing with dynamic aspects, and those dealing with research evaluation, among others. Rhythm indicators basically connect the first two types.

Fundamental and derived properties of rhythm sequences have been explored in (Liang & Rousseau, 2007; Egghe, Liang & Rousseau, 2008). Moreover, the idea of a journal's input rhythm has been proposed and applied to journals such as *Science* (Liang et al., 2006; Liang & Rousseau, 2010).

In previous studies, $O_i$ (observed values) and $E_i$ (expected values) were calculated based on the same p-c matrix of the actor under study, which might be a research group, a journal, an institute, a country, a field, or even a prolific scientist. The calculation of the expected values $E_i$ does not increase or decrease the total number of citations in the study. It is just a redistribution of observation-based citations over publication years. Therefore, $R_i$ is an indicator describing the internal rhythm of an – isolated - actor.

What happens if $O_i$ and $E_i$ are calculated based on two different p-c matrices? This idea implies breaking the actor's own scope and comparing its observation-based citations with its expectation-based citations, which is related to "the outer world". The results of this operation can form another kind of R-sequence, namely the external rhythm of an actor. In this paper, we will define and create the R-sequence of an actor's external rhythm.

Furthermore, since the external rhythm connects an actor to its environment and makes a comparison between the actor and the environment possible, we may ask if this construction can be used to construct a new citation-based indicator for the study of science or for research evaluation. We will discuss this question and provide examples.

## The R-sequence of the internal and the external rhythm of an actor

### *Creating an R-sequence for the internal rhythm of an actor based on its p-c matrix*

To explain how to create the external rhythm of an actor, we first review the method of creating an R-sequence of the internal rhythm. We note that we always assume that a fixed database is used. The starting point is a general p-c matrix of the actor under study. Table 1 provides such an example. Here, $P_i$ is the number of publications published in the year $i$ (in general: time interval $i$). The notation $C_{ij}$ stands for the number of citations received in the year $j$ by items published in the year $i$, $j \geq i$. In this work, we always use whole counting for publications and citations, but we refer the reader to the discussion about this point further on.

Table 1. General p-c matrix

| | | | citing year $j$ and received citations $C_{ij}$ | | | | | | | | |
|---|---|---|---|---|---|---|---|---|---|---|---|
| | | | 1 | 2 | 3 | 4 | 5 | 6 | 7 | 8 | 9 |
| $i$ and number of publications | 1 | $P_1$ | $C_{11}$ | $C_{12}$ | $C_{13}$ | $C_{14}$ | $C_{15}$ | $C_{16}$ | $C_{17}$ | $C_{18}$ | $C_{19}$ |
| | 2 | $P_2$ | | $C_{22}$ | $C_{23}$ | $C_{24}$ | $C_{25}$ | $C_{26}$ | $C_{27}$ | $C_{28}$ | $C_{29}$ |
| | 3 | $P_3$ | | | $C_{33}$ | $C_{34}$ | $C_{35}$ | $C_{36}$ | $C_{37}$ | $C_{38}$ | $C_{39}$ |

| | | | | | | | | | | |
|---|---|---|---|---|---|---|---|---|---|---|
| 4 | $P_4$ | | | | $C_{44}$ | $C_{45}$ | $C_{46}$ | $C_{47}$ | $C_{48}$ | $C_{49}$ |
| 5 | $P_5$ | | | | | $C_{55}$ | $C_{56}$ | $C_{57}$ | $C_{58}$ | $C_{59}$ |
| 6 | $P_6$ | | | | | | $C_{66}$ | $C_{67}$ | $C_{68}$ | $C_{69}$ |
| 7 | $P_7$ | | | | | | | $C_{77}$ | $C_{78}$ | $C_{79}$ |
| 8 | $P_8$ | | | | | | | | $C_{88}$ | $C_{89}$ |
| 9 | $P_9$ | | | | | | | | | $C_{99}$ |

In general, we will consider a period of length $n$ (in Table 1, $n$ = 9). The symbols $O_i$ and $E_i$ are defined as follows:

$$O_i = \sum_{j=i}^{n} C_{ij} \qquad i = 1, \ldots, n \tag{1}$$

$$E_i = P_i\left(\sum_{k=1}^{n-i=1} C_k\right) \quad , i = 1, \ldots, n \tag{2}$$

$O_i$, i =1, …, n in formula (1) denotes the number of citations received by the publications of the year i within the timeframe of the p-c matrix. $C_k$ is defined in formula (3) and denotes the average number of citations per paper in the $k^{th}$ year after its publication, $k = 1$ to $n$, where $k = 1$ refers to a paper's publication year. $C_k$ is the key measure in the construction of the R-sequence. It is defined in formula (3) and is equal to the average number of citations in the p-c matrix, received k-1 years after publication, where year 1 is the

publication year:

$$C_k = \frac{\sum_{j=1}^{n-k+1} C_{j,j+k-1}}{\sum_{j=1}^{n-k+1} P_j} \qquad (3)$$

Note that $C_n = \frac{C_{1,n}}{P_1}$. Consequently, $E_i$ is equal to the number of publications in the year i, multiplied by the sum of the average number of citations in the p-c matrix, received during the period between the publication and the end of the citation window. The indicator R is defined as a time series of ratios

$$R_i = \frac{O_i}{E_i}, \; i = 1, \ldots, n.$$

The numerator of each ratio, denoted as $O_i$, is an actual (observed) citation value, while the denominator, denoted as $E_i$, is an expected citation value. The sequence $R_1, R_2, \ldots, R_n$ is called the R-sequence, a time series of the actor's citation performance. It is interpreted as an aspect of the internal rhythm of the actor.

### *Creating an R-sequence of the external rhythm based on two p-c matrices*

Now we approach the problem of calculating $O_i$ and $E_i$ based on two different p-c matrices. Assume that $A$ is a collective, consisting of m disjoint, non-empty parts: $A = \bigcup_{q=1}^{m} \boldsymbol{B_q}$. Mathematically speaking, the sets $B_q$ form a partition of $A$. In practical applications, $A$ can be a field as represented by articles published in journals $B_q$, a country covering many regions $B_q$, a university composed of different departments $B_q$, and even the whole world of science, considered as a collective of (disjoint) fields. Corresponding to the collective A, the $B_q$ ($q = 1, 2, \ldots, m$) are called the constituents. We want to study $B_q$ with respect to A, of course, without $B_q$ itself, i.e., $A \setminus B_q$ denoted as $A_q$. If we consider the same time span for the publications as for citations, say $n$ years, we can calculate the $P_i$ and $C_{ij}$ data and establish the p-c matrix of $A_q$. Then we can calculate $C_k$ using formula (3) based on the p-c matrix $A_q$. Results are denoted as $\left(\boldsymbol{C}_{A_q}\right)_{\boldsymbol{k}}$ $k = 1, \ldots, n,$ where $k = 1$ refers to the publication year. Recall that $\left(\boldsymbol{C}_{A_q}\right)_{\boldsymbol{k}}$ represents the average number of citations per paper received in the $k^{th}$ year after its publication. Here, the average is taken over the collective $A_q$. Based on the same citation window and the same citation database, we can further construct p-c matrices for all constituents $B_q$ ($q=1, 2, \ldots, m$). The values $C_k$ derived from $B_q$ are denoted as $\left(\boldsymbol{C}_{B_q}\right)_{\boldsymbol{k}}$.

The next step is constructing $R_i$ with the $O_i$ and $E_i$ calculated based on different p-c matrices. In this construction, we will use the following symbols: the notations $\left(\boldsymbol{P}_{B_q}\right)_{\boldsymbol{i}}$ and $\left(\boldsymbol{C}_{B_q}\right)_{\boldsymbol{ij}}$ are used to illustrate that these numbers $P_i$ and $C_{ij}$ come from the p-c matrix $B_q$. Similarly, $\left(\boldsymbol{P}_{A_q}\right)_{\boldsymbol{i}}$ and $\left(\boldsymbol{C}_{A_q}\right)_{\boldsymbol{ij}}$ are obtained from the p-c matrix $A_q$.

Next, we explain how to construct $R_i$.

First, we calculate $O_i$ (i refers to a publication year) according to formula (1) based on the p-c matrix $B_q$, $q \in \{1, 2, \ldots, m\}$, and denote the results as $\left(\boldsymbol{O}_{\boldsymbol{B_q}}\right)_{\boldsymbol{i}}$:

$$\left(\boldsymbol{O}_{\boldsymbol{B_q}}\right)_{\boldsymbol{i}} = \sum_{\boldsymbol{j}=\boldsymbol{i}}^{\boldsymbol{n}} \left(\boldsymbol{C}_{\boldsymbol{B_q}}\right)_{\boldsymbol{ij}} \tag{4}$$

Secondly, according to formula (2), we calculate $E_i$ with $P_i$ coming from the p-c matrix $B_q$ and $C_k$ from the p-c matrix $A_q$. We denote the results as $\left(\boldsymbol{E}_{\boldsymbol{B_q};\boldsymbol{A_q}}\right)_{\boldsymbol{i}}$. Then,

$$\left(\boldsymbol{E}_{\boldsymbol{B_q};\boldsymbol{A_q}}\right)_{\boldsymbol{i}} = (\boldsymbol{P}_{\boldsymbol{B_q}})_{\boldsymbol{i}} \sum_{\boldsymbol{k}=\boldsymbol{1}}^{\boldsymbol{n}-\boldsymbol{i}+\boldsymbol{1}} (\boldsymbol{C}_{\boldsymbol{A_q}})_{\boldsymbol{k}} \tag{5}$$

Here, all $\left(\boldsymbol{C}_{\boldsymbol{A_q}}\right)_{\boldsymbol{k}}$ are calculated according to formula (3) based on the data in the p-c matrix $A_q$.

Thirdly, we define the ratio of $\left(\boldsymbol{O}_{\boldsymbol{B_q}}\right)_{\boldsymbol{i}}$ to $\left(\boldsymbol{E}_{\boldsymbol{B_q};\boldsymbol{A_q}}\right)_{\boldsymbol{i}}$ as $\left(\boldsymbol{R}_{\boldsymbol{B_q};\boldsymbol{A_q}}\right)_{\boldsymbol{i}}$

$$\left(\boldsymbol{R}_{\boldsymbol{B_q};\boldsymbol{A_q}}\right)_{\boldsymbol{i}} = \left(\boldsymbol{O}_{\boldsymbol{B_q}}\right)_{\boldsymbol{i}} / \left(\boldsymbol{E}_{\boldsymbol{B_q};\boldsymbol{A_q}}\right)_{\boldsymbol{i}} \tag{6}$$

Obviously, in formula (6) $\left(\boldsymbol{O}_{\boldsymbol{B_q}}\right)_{\boldsymbol{i}}$ and $\left(\boldsymbol{E}_{\boldsymbol{B_q};\boldsymbol{A_q}}\right)_{\boldsymbol{i}}$ come from different p-c matrices. For actor $B_q$, $\left(\boldsymbol{O}_{\boldsymbol{B_q}}\right)_{\boldsymbol{i}}$ are observed citation values, $\left(\boldsymbol{E}_{\boldsymbol{B_q};\boldsymbol{A_q}}\right)_{\boldsymbol{i}}$ are expected citation values.

In Table 2 we list the possible $R_i$ with $O_i$ and $E_i$ calculated based on p-c matrices $A_q$ or $B_q$. The pairs of p-c matrices for $O_i$ and $E_i$ are: ($B_q$ – $B_q$), ($B_q$ – $A_q$), ($A_q$ – $B_q$), and ($A_q$ – $A_q$), always taken in that order.

Table 2. $R_i$ with $O_i$ and $E_i$ calculated based on two different p-c

matrices

| | $B_q$ (*q=1, 2, … , m*) | $A_q$ (*q=1, 2, … , m*) |
|---|---|---|
| $B_q$ (*q=1, 2, … , m*) | $\left(\boldsymbol{O}_{B_q}\right)_i = \sum_{j=i}^{n}\left(\boldsymbol{C}_{B_q}\right)_{ij}$ <br> $\left(\boldsymbol{E}_{B_q}\right)_i = \left(\boldsymbol{P}_{B_q}\right)_i \sum_{k=1}^{n-i+1}\left(\boldsymbol{C}_{B_q}\right)_k$ <br> $\left(\boldsymbol{R}_{B_q}\right)_i = \left(\boldsymbol{O}_{B_q}\right)_i / \left(\boldsymbol{E}_{B_q}\right)_i$ | $\left(\boldsymbol{O}_{B_q}\right)_i = \sum_{j=i}^{n}\left(\boldsymbol{C}_{B_q}\right)_{ij}$ <br> $\left(\boldsymbol{E}_{B_q;A_q}\right)_i = (\boldsymbol{P}_{B_q})_i \sum_{k=1}^{n-i+1}(\boldsymbol{C}_{A_q})_k$ <br> $\left(\boldsymbol{R}_{B_q;A_q}\right)_i = \left(\boldsymbol{O}_{B_q}\right)_i / \left(\boldsymbol{E}_{B_q;A_q}\right)_i$ |
| $A_q$ (*q=1, 2, … , m*) | $\left(\boldsymbol{O}_{A_q}\right)_i = \sum_{j=i}^{n}\left(\boldsymbol{C}_{A_q}\right)_{ij}$ <br> $(\boldsymbol{E}_{A_q;B_q})_i = (\boldsymbol{P}_{A_q})_i \sum_{k=1}^{n-i+1}(\boldsymbol{C}_{B_q})_k$ <br> $\left(\boldsymbol{R}_{A_q;B_q}\right)_i = \left(\boldsymbol{O}_{A_q}\right)_i / \left(\boldsymbol{E}_{A_q;B_q}\right)_i$ | $\left(\boldsymbol{O}_{A_q}\right)_i = \sum_{j=i}^{n}\left(\boldsymbol{C}_{A_q}\right)_{ij}$ <br> $\left(\boldsymbol{E}_{A_q}\right)_i = (\boldsymbol{P}_{A_q})_i \sum_{k=1}^{n-i+1}(\boldsymbol{C}_{A_q})_k$ <br> $\left(\boldsymbol{R}_{A_q}\right)_i = \left(\boldsymbol{O}_{A_q}\right)_i / \left(\boldsymbol{E}_{A_q}\right)_i$ |

We briefly discuss all R-sequences in Table 2, and, for the reader's convenience, determine which reflect an internal rhythm and which reflect an external one.

- In constructing the sequence $\left(\boldsymbol{R}_{B_q}\right)_i$ the numerator $\left(\boldsymbol{O}_{B_q}\right)_i$ and the denominator $\left(\boldsymbol{E}_{B_q}\right)_i$ are calculated based on the same p-c matrix $B_q$, hence the sequence $\left(\boldsymbol{R}_{B_q}\right)_i$ reflects the internal rhythm of $B_q$.
- In constructing the sequence $\left(\boldsymbol{R}_{A_q}\right)_i$ or $(\boldsymbol{R}_A)_i$ numerators and denominators are based on the same p-c matrix $A_q$ *(or A)*, hence the corresponding R-sequence reflects the internal rhythm of $A_q$ (or *A)*.

- In constructing the sequence $\left(\boldsymbol{R}_{B_q;\,A_q}\right)_i$ the numerator $\left(\boldsymbol{O}_{B_q}\right)_i$ is calculated based on the p-c matrix $B_q$, while the $(\boldsymbol{C}_{A_q})_k$ , the main ingredient of the denominator $\left(\boldsymbol{E}_{B_q;\,A_q}\right)_i$ , is calculated based on the p-c matrix $A_q$. Hence, the sequence $\left(\boldsymbol{R}_{B_q;\,A_q}\right)_i$ is created based on two different matrices $B_q$ and $A_q$, reflecting an external rhythm of actor $B_q$. Here, the collective $A_q$ is actor $B_q$'s "outer world". This is a comparison between an actor and the collective to which the actor belongs.
- The sequence $\left(\boldsymbol{R}_{A_q;\,B_q}\right)_i$ does not seem meaningful unless we want to compare the average level of a collective with a single $B_q$.

*A summary indicator*

While in the internal case $\sum_i \boldsymbol{O}_i = \sum_i \boldsymbol{E}_i$ is an invariant of the system (Liang & Rousseau, 2007), in the external case $\sum_i \boldsymbol{O}_i$ is not necessarily equal to $\sum_i \boldsymbol{E}_i$. Hence, $\boldsymbol{I}_1 = \frac{\sum_i \boldsymbol{O}_i}{\sum_i \boldsymbol{E}_i}$ can be used as a summary indicator. Alternatively, the average of ratios indicator $I_2 = \frac{1}{n} \sum_{i=1}^{n} \left(\frac{O_i}{E_i}\right)$ can be used. We note that, also in the case of an internal rhythm, the average of ratios $I_2$ can be used as a summary indicator.

**Can the external rhythm of an actor be used in evaluation and science of science studies?**

Can creating the external rhythm be translated into a tool for research studies, maybe even for research evaluation? We will discuss this problem for two situations:

1) comparing the citation performance of a single actor $B_q$ with the average citation performance of the collective $A$, to which actor $B_q$ belongs;

2) comparing the citation performance of a single actor with another single actor in the same collective $A$.

For situation 1), we consider $B_q$'s external rhythm sequence ($\left(\boldsymbol{R}_{\boldsymbol{B}_q;\,\boldsymbol{A}_q}\right)_{\boldsymbol{i}} = \left(\boldsymbol{O}_{\boldsymbol{B}_q}\right)_{\boldsymbol{i}} / \left(\boldsymbol{E}_{\boldsymbol{B}_q;\,\boldsymbol{A}_q}\right)_{\boldsymbol{i}}$), $i$ = 1, …, $n$. The sequence $\left(\boldsymbol{R}_{\boldsymbol{B}_q;\,\boldsymbol{A}_q}\right)_{\boldsymbol{i}}$ is the result of comparing the citation performance of actor $B_q$ to the average citation performance of the collective $A$ (excluding $B_q$). This result may reflect the difference between an ordinary actor and the average level of the collective $A$. If the ratio $\left(\boldsymbol{R}_{\boldsymbol{B}_q;\,\boldsymbol{A}_q}\right)_{\boldsymbol{i}}$ is bigger than 1, this means that actor $B_q$'s actual performance in the year $i$ is better than the average level of the collective A in the same year. Conversely, if the ratio $\left(\boldsymbol{R}_{\boldsymbol{B}_q;\,\boldsymbol{A}_q}\right)_{\boldsymbol{i}}$ is smaller than 1, this means that actor $B_q$'s actual performance in the year $i$ is inferior to the average level of collective A (without $B_q$) in the same year. In case the ratio $\left(\boldsymbol{R}_{\boldsymbol{B}_q;\,\boldsymbol{A}_q}\right)_{\boldsymbol{i}}$ is about 1, actor $B_q$'s citation performance in the year $i$ is (almost) equal to collective $A_q$'s average level. Thus, for *i=1, 2, ...,n* the sequence $\left(\boldsymbol{R}_{\boldsymbol{B}_q;\,\boldsymbol{A}_q}\right)_{\boldsymbol{i}}$ may go up and down, i.e., fluctuate, reflecting the changes in the actor's citation performance over the years. For this reason an external rhythm sequence $\left(\boldsymbol{R}_{\boldsymbol{B}_q;\,\boldsymbol{A}_q}\right)_{\boldsymbol{i}}$ can be used in science studies.

Situation (2) is slightly more complicated. In section 2.2 we assumed that *A* is a collective, $A = \cup_{q=1}^{m} \boldsymbol{B}_{\boldsymbol{q}}$. Suppose now that $B_u$ and $B_v$ are two actors in *A*, $u,v \in \{1, 2, \ldots, m\}$. Then, to avoid comparing $B_u$ ($B_v$) partly to itself, and to avoid comparing $B_u$ and ($B_v$) to different sets ($A \setminus B_u$ and $A \setminus B_v$) we have to use $\boldsymbol{A}_{\boldsymbol{u},\boldsymbol{v}} = \boldsymbol{A} \setminus (\boldsymbol{B}_{\boldsymbol{u}} \cup \boldsymbol{B}_{\boldsymbol{v}})$ to determine the sequences $\left(\boldsymbol{R}_{\boldsymbol{B}_u;\,\boldsymbol{A}_{u,v}}\right)_{\boldsymbol{i}}$ and $\left(\boldsymbol{R}_{\boldsymbol{B}_v;\,\boldsymbol{A}_{u,v}}\right)_{\boldsymbol{i}}$ and make a comparison between them. If, for a fixed year *i*, $\left(\boldsymbol{R}_{\boldsymbol{B}_u;\,\boldsymbol{A}_{u,v}}\right)_{\boldsymbol{i}} > \left(\boldsymbol{R}_{\boldsymbol{B}_v;\,\boldsymbol{A}_{u,v}}\right)_{\boldsymbol{i}}$ $B_u$'s citation performance is better than that of $B_v$ in the year *i*.

**Some examples**

We will provide an example of case (1) as well as case (2). Let collective *A* consist of all the publications of article or review type, published in the journal *Scientometrics,* in short SCIM*,* during the period 2015 to 2024. For the first example, $B_q$ consists of all the SCIM publications (2015-2024) with the corresponding author from China. More precisely, those publications for which the first affiliation of the first corresponding author is in China. For the second example, $B_u$ consists of all SCIM publications, again published during the period (2015-2024), with the corresponding author from Brazil, while $B_v$ consists of all SCIM publications with the corresponding author from the Netherlands.

The data are retrieved from Clarivate's Web of Science, core collection, in October 2025.

***Case 1: Comparing one country (China) with all others***

Table 3 gives the data for China in SCIM. The first column refers to the publication year (FPY in the WoS). The second column (Publ) shows the number of items published in the journal SCIM for which the corresponding author has a Chinese affiliation. The other cells refer to received citations in different years. The sums of all received citations are the O-values of a particular year, shown in the last column.

Table 3.The p-c matrix $B_q$ (China)

| Year | Publ | 2015 | 2016 | 2017 | 2018 | 2019 | 2020 | 2021 | 2022 | 2023 | 2024 | O-values |
|---|---|---|---|---|---|---|---|---|---|---|---|---|
| 2015 | 74 | 23 | 104 | 222 | 221 | 247 | 305 | 296 | 274 | 257 | 200 | 2149 |
| 2016 | 48 | | 11 | 80 | 133 | 149 | 183 | 167 | 175 | 160 | 147 | 1205 |
| 2017 | 67 | | | 36 | 125 | 258 | 219 | 285 | 245 | 187 | 182 | 1537 |
| 2018 | 66 | | | | 41 | 183 | 255 | 332 | 329 | 273 | 295 | 1708 |
| 2019 | 60 | | | | | 44 | 158 | 214 | 249 | 205 | 220 | 1090 |
| 2020 | 91 | | | | | | 98 | 365 | 538 | 509 | 614 | 2124 |
| 2021 | 86 | | | | | | | 43 | 227 | 278 | 271 | 819 |
| 2022 | 86 | | | | | | | | 105 | 240 | 361 | 706 |
| 2023 | 80 | | | | | | | | | 63 | 237 | 300 |
| 2024 | 101 | | | | | | | | | | 84 | 84 |

Based on Table 3, we calculate the internal rhythm for China in SCIM, see Table 4. The summary indicators for China are $I_1 = 1$, and $I_2 = 1.036$.

Table 4. Calculations leading to the internal R-sequence of China

| $O_i$ | | $C_k$ | | $E_i$ | | $R_i$ | |
|---|---|---|---|---|---|---|---|
| $O_{2015}$ | 2149 | $C_1$ | 0.722 | $E_{2015}$ | 2415.864 | $R_{2015}$ | 0.890 |
| $O_{2016}$ | 1205 | $C_2$ | 2.612 | $E_{2016}$ | 1437.317 | $R_{2016}$ | 0.838 |
| $O_{2017}$ | 1537 | $C_3$ | 3.908 | $E_{2017}$ | 1784.387 | $R_{2017}$ | 0.861 |
| $O_{2018}$ | 1708 | $C_4$ | 3.963 | $E_{2018}$ | 1542.643 | $R_{2018}$ | 1.107 |
| $O_{2019}$ | 1090 | $C_5$ | 4.589 | $E_{2019}$ | 1178.167 | $R_{2019}$ | 0.925 |
| $O_{2020}$ | 2124 | $C_6$ | 3.841 | $E_{2020}$ | 1437.332 | $R_{2020}$ | 1.478 |
| $O_{2021}$ | 819 | $C_7$ | 3.737 | $E_{2021}$ | 963.732 | $R_{2021}$ | 0.850 |
| $O_{2022}$ | 706 | $C_8$ | 3.259 | $E_{2022}$ | 622.878 | $R_{2022}$ | 1.133 |
| $O_{2023}$ | 300 | $C_9$ | 3.311 | $E_{2023}$ | 266.757 | $R_{2023}$ | 1.125 |
| $O_{2024}$ | 84 | $C_{10}$ | 2.703 | $E_{2024}$ | 72.922 | $R_{2024}$ | 1.152 |
| sum | 11722 | | | | 11722 | | |

Table 5. The p-c matrix $A \setminus B_q$ (SCIM without China)

| Year | Publ | 2015 | 2016 | 2017 | 2018 | 2019 | 2020 | 2021 | 2022 | 2023 | 2024 | O-values |
|---|---|---|---|---|---|---|---|---|---|---|---|---|
| 2015 | 275 | 157 | 581 | 694 | 762 | 891 | 992 | 1092 | 1213 | 1061 | 1105 | 8548 |
| 2016 | 301 | | 180 | 752 | 1030 | 1403 | 1449 | 1508 | 1635 | 1540 | 1630 | 11127 |
| 2017 | 305 | | | 213 | 753 | 1176 | 1165 | 1314 | 1335 | 1343 | 1203 | 8502 |
| 2018 | 307 | | | | 176 | 835 | 1007 | 1249 | 1208 | 1091 | 1049 | 6615 |
| 2019 | 215 | | | | | 132 | 633 | 941 | 978 | 902 | 859 | 4445 |
| 2020 | 303 | | | | | | 228 | 951 | 1115 | 958 | 969 | 4221 |
| 2021 | 309 | | | | | | | 367 | 1154 | 1443 | 1638 | 4602 |
| 2022 | 242 | | | | | | | | 220 | 709 | 933 | 1862 |
| 2023 | 191 | | | | | | | | | 170 | 521 | 691 |
| 2024 | 214 | | | | | | | | | | 231 | 231 |

Based on this matrix (Table 5), we calculate the internal rhythm for SCIM without China.

Table 6. Calculations leading to the internal R-sequence for SCIM without China

| $O_i$ | | $C_k$ | | $E_i$ (rounded) | | $R_i$ | |
|---|---|---|---|---|---|---|---|
| $O_{2015}$ | 8548 | $C_1$ | 0.779 | $E_{2015}$ | 10146 | $R_{2015}$ | 0.843 |
| $O_{2016}$ | 11127 | $C_2$ | 2.814 | $E_{2016}$ | 9895 | $R_{2016}$ | 1.124 |
| $O_{2017}$ | 8502 | $C_3$ | 3.695 | $E_{2017}$ | 8602 | $R_{2017}$ | 0.988 |
| $O_{2018}$ | 6615 | $C_4$ | 4.046 | $E_{2018}$ | 7280 | $R_{2018}$ | 0.909 |
| $O_{2019}$ | 4445 | $C_5$ | 3.947 | $E_{2019}$ | 4172 | $R_{2019}$ | 1.065 |
| $O_{2020}$ | 4221 | $C_6$ | 4.123 | $E_{2020}$ | 4630 | $R_{2020}$ | 0.912 |
| $O_{2021}$ | 4602 | $C_7$ | 4.309 | $E_{2021}$ | 3502 | $R_{2021}$ | 1.314 |
| $O_{2022}$ | 1862 | $C_8$ | 4.490 | $E_{2022}$ | 1764 | $R_{2022}$ | 1.056 |
| $O_{2023}$ | 691 | $C_9$ | 4.672 | $E_{2023}$ | 686 | $R_{2023}$ | 1.007 |
| $O_{2024}$ | 231 | $C_{10}$ | 4.018 | $E_{2024}$ | 167 | $R_{2024}$ | 1.385 |
| sum | 50844 | | | | 50843 | | |

From this, we see that for SCIM without China, $I_1$=1 and $I_2$ = 1.060.

Now we come to the main calculations, namely comparing China with the rest of the world (SCIM). For this, we compare the observed values for China with the expected values derived from the $C_k$-values of the world without China (in SCIM). Data are shown in Table 7. The summary indicators for China with respect to the world (in SCIM) are: $I_1$= 0.9542, and $I_2$=0.9997 (≈ 1).

Table 7. Calculations leading to the R-sequence of China in SCIM.

| $P_i$ China | | $O_i$ China | | $C_k$ World | | $E_i$ World | | $R_i$ external | |
|---|---|---|---|---|---|---|---|---|---|
| $P_{2015}$ | 74 | $O_{2015}$ | 2149 | $C_1$ | 0.779 | $E_{2015}$ | 2730.114 | $R_{2015}$ | 0.787 |
| $P_{2016}$ | 48 | $O_{2016}$ | 1205 | $C_2$ | 2.814 | $E_{2016}$ | 1578.012 | $R_{2016}$ | 0.764 |
| $P_{2017}$ | 67 | $O_{2017}$ | 1537 | $C_3$ | 3.695 | $E_{2017}$ | 1889.626 | $R_{2017}$ | 0.813 |
| $P_{2018}$ | 66 | $O_{2018}$ | 1708 | $C_4$ | 4.046 | $E_{2018}$ | 1565.059 | $R_{2018}$ | 1.091 |
| $P_{2019}$ | 60 | $O_{2019}$ | 1090 | $C_5$ | 3.947 | $E_{2019}$ | 1164.246 | $R_{2019}$ | 0.936 |
| $P_{2020}$ | 91 | $O_{2020}$ | 2124 | $C_6$ | 4.123 | $E_{2020}$ | 1390.552 | $R_{2020}$ | 1.527 |
| $P_{2021}$ | 86 | $O_{2021}$ | 819 | $C_7$ | 4.309 | $E_{2021}$ | 974.735 | $R_{2021}$ | 0.840 |
| $P_{2022}$ | 86 | $O_{2022}$ | 706 | $C_8$ | 4.490 | $E_{2022}$ | 626.766 | $R_{2022}$ | 1.126 |
| $P_{2023}$ | 80 | $O_{2023}$ | 300 | $C_9$ | 4.672 | $E_{2023}$ | 287.460 | $R_{2023}$ | 1.044 |
| $P_{2024}$ | 101 | $O_{2024}$ | 84 | $C_{10}$ | 4.018 | $E_{2024}$ | 78.690 | $R_{2024}$ | 1.067 |
| sum | | | 11722 | | | | 12285.26 | | |

Fig. 1 shows the external R-sequence calculated in Table 7. The summary indicator for China with respect to the world (in SCIM) is equal to 1 (actually, 0.9997).

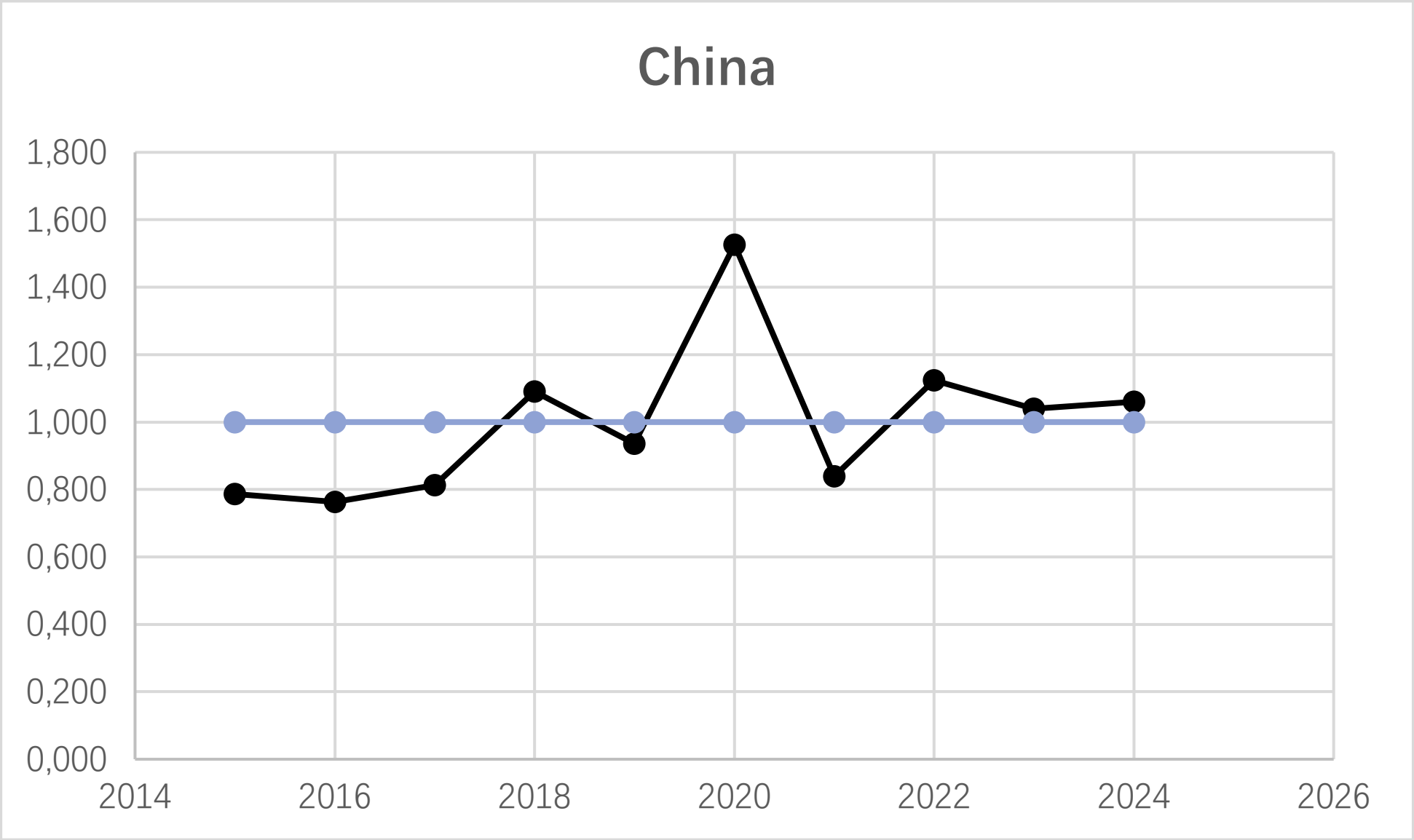

Fig. 1 China's external rhythm sequence

The peak in 2020 (Fig. 1) corresponds to an article by Zhu and Liu (2020), which had already received 837 citations in October 2025.

### *Case 2: Comparing Brazil to the Netherlands in SCIM*

Table 8.The p-c matrix $B_u$ (Brazil)

| Year | Publ | 2015 | 2016 | 2017 | 2018 | 2019 | 2020 | 2021 | 2022 | 2023 | 2024 | O-values |
|---|---|---|---|---|---|---|---|---|---|---|---|---|
| 2015 | 14 | 11 | 11 | 28 | 19 | 50 | 58 | 59 | 56 | 48 | 48 | 388 |
| 2016 | 17 | | 7 | 24 | 38 | 56 | 46 | 33 | 41 | 28 | 16 | 289 |
| 2017 | 18 | | | 4 | 39 | 65 | 74 | 131 | 104 | 74 | 68 | 559 |
| 2018 | 15 | | | | 4 | 45 | 57 | 69 | 76 | 80 | 79 | 410 |
| 2019 | 11 | | | | | 4 | 17 | 30 | 39 | 40 | 20 | 150 |
| 2020 | 17 | | | | | | 6 | 42 | 49 | 45 | 36 | 178 |
| 2021 | 19 | | | | | | | 18 | 50 | 60 | 56 | 184 |
| 2022 | 14 | | | | | | | | 5 | 37 | 54 | 96 |
| 2023 | 11 | | | | | | | | | 8 | 29 | 37 |
| 2024 | 5 | | | | | | | | | | 4 | 4 |

Table 9. Calculations leading to the internal R-sequence of Brazil

| $O_i$ | | $C_k$ | | $E_i$ | | $R_i$ | |
|---|---|---|---|---|---|---|---|
| $O_{2015}$ | 388 | $C_1$ | 0.504 | $E_{2015}$ | 413.556 | $R_{2015}$ | 0.938 |
| $O_{2016}$ | 289 | $C_2$ | 2.162 | $E_{2016}$ | 443.889 | $R_{2016}$ | 0.651 |
| $O_{2017}$ | 559 | $C_3$ | 3.048 | $E_{2017}$ | 432.839 | $R_{2017}$ | 1.291 |
| $O_{2018}$ | 410 | $C_4$ | 3.225 | $E_{2018}$ | 314.168 | $R_{2018}$ | 1.305 |
| $O_{2019}$ | 150 | $C_5$ | 4.120 | $E_{2019}$ | 186.906 | $R_{2019}$ | 0.803 |
| $O_{2020}$ | 178 | $C_6$ | 3.933 | $E_{2020}$ | 221.988 | $R_{2020}$ | 0.802 |
| $O_{2021}$ | 184 | $C_7$ | 3.953 | $E_{2021}$ | 169.832 | $R_{2021}$ | 1.083 |
| $O_{2022}$ | 96 | $C_8$ | 3.102 | $E_{2022}$ | 79.986 | $R_{2022}$ | 1.200 |
| $O_{2023}$ | 37 | $C_9$ | 2.065 | $E_{2023}$ | 29.318 | $R_{2023}$ | 1.262 |
| $O_{2024}$ | 4 | $C_{10}$ | 3.429 | $E_{2024}$ | 2.518 | $R_{2024}$ | 1.589 |
| sum | 2295 | | | | 2295 | | |

We obtain: $I_1$=1 (by definition), $I_2$ (Brazil) = 1.092. Next, we show similar data and calculations for the Netherlands.

Table 10. The p-c matrix $B_v$ (The Netherlands)

| Year | Publ | 2015 | 2016 | 2017 | 2018 | 2019 | 2020 | 2021 | 2022 | 2023 | 2024 | O-values |
|---|---|---|---|---|---|---|---|---|---|---|---|---|
| 2015 | 9 | 6 | 29 | 23 | 32 | 18 | 36 | 25 | 32 | 24 | 21 | 246 |
| 2016 | 13 | | 15 | 58 | 80 | 92 | 76 | 80 | 95 | 97 | 97 | 690 |
| 2017 | 10 | | | 27 | 60 | 133 | 155 | 210 | 294 | 419 | 401 | 1699 |
| 2018 | 7 | | | | 7 | 30 | 24 | 42 | 49 | 39 | 42 | 233 |
| 2019 | 5 | | | | | 1 | 32 | 31 | 36 | 29 | 26 | 155 |
| 2020 | 5 | | | | | | 10 | 21 | 30 | 29 | 31 | 121 |
| 2021 | 6 | | | | | | | 7 | 23 | 33 | 35 | 98 |
| 2022 | 1 | | | | | | | | 0 | 1 | 1 | 2 |
| 2023 | 5 | | | | | | | | | 7 | 30 | 37 |
| 2024 | 9 | | | | | | | | | | 9 | 9 |

Based on Table 10, we calculate the internal rhythm of the Netherlands in SCIM, see Table 11.

Table 11. Calculations leading to the internal rhythm of the Netherlands in SCIM.

| $O_i$ | | $C_k$ | | $E_i$ | | $R_i$ | |
|---|---|---|---|---|---|---|---|
| $O_{2015}$ | 246 | $C_1$ | 1.271 | $E_{2015}$ | 705.945 | $R_{2015}$ | 0.348 |
| $O_{2016}$ | 690 | $C_2$ | 4.656 | $E_{2016}$ | 989.364 | $R_{2016}$ | 0.697 |
| $O_{2017}$ | 1699 | $C_3$ | 6.339 | $E_{2017}$ | 706.050 | $R_{2017}$ | 2.406 |
| $O_{2018}$ | 233 | $C_4$ | 7.655 | $E_{2018}$ | 378.297 | $R_{2018}$ | 0.616 |
| $O_{2019}$ | 155 | $C_5$ | 8.429 | $E_{2019}$ | 195.725 | $R_{2019}$ | 0.792 |
| $O_{2020}$ | 121 | $C_6$ | 10.795 | $E_{2020}$ | 141.748 | $R_{2020}$ | 0.854 |
| $O_{2021}$ | 98 | $C_7$ | 14.897 | $E_{2021}$ | 119.526 | $R_{2021}$ | 0.820 |
| $O_{2022}$ | 2 | $C_8$ | 16.563 | $E_{2022}$ | 12.266 | $R_{2022}$ | 0.163 |
| $O_{2023}$ | 37 | $C_9$ | 5.500 | $E_{2023}$ | 29.636 | $R_{2023}$ | 1.248 |
| $O_{2024}$ | 9 | $C_{10}$ | 2.333 | $E_{2024}$ | 11.443 | $R_{2024}$ | 0.787 |
| sum | 3290 | | | | 3290 | | |

For the Netherlands, we find: $I_1 = 1$, $I_2$(Netherlands) = 0.873. Finally, we show in Table 12 the data and calculations for SCIM without the union of Brazil and the Netherlands.

Table 12. Calculations leading to the internal R-sequence for SCIM without the union of Brazil and the Netherlands.

| Year | Publ | 2015 | 2016 | 2017 | 2018 | 2019 | 2020 | 2021 | 2022 | 2023 | 2024 | O-values |
|---|---|---|---|---|---|---|---|---|---|---|---|---|
| 2015 | 326 | 163 | 645 | 865 | 932 | 1070 | 1203 | 1304 | 1399 | 1246 | 1236 | 10063 |
| 2016 | 319 | | 169 | 750 | 1045 | 1404 | 1510 | 1562 | 1674 | 1575 | 1664 | 11353 |
| 2017 | 344 | | | 218 | 779 | 1236 | 1155 | 1258 | 1182 | 1037 | 916 | 7781 |
| 2018 | 351 | | | | 206 | 943 | 1181 | 1470 | 1412 | 1245 | 1223 | 7680 |
| 2019 | 259 | | | | | 171 | 742 | 1094 | 1152 | 1038 | 1033 | 5230 |
| 2020 | 372 | | | | | | 310 | 1253 | 1574 | 1393 | 1516 | 6046 |
| 2021 | 370 | | | | | | | 385 | 1308 | 1628 | 1818 | 5139 |
| 2022 | 313 | | | | | | | | 320 | 911 | 1239 | 2470 |
| 2023 | 255 | | | | | | | | | 218 | 699 | 917 |
| 2024 | 301 | | | | | | | | | | 302 | 302 |

Table 13. Calculations leading to the internal rhythm of SCIM without the union of Brazil and the Netherlands.

| $O_i$ | | $C_k$ | | $E_i$ | | $R_i$ | |
|---|---|---|---|---|---|---|---|
| $O_{2015}$ | 10063 | $C_1$ | 0.771 | $E_{2015}$ | 11487.009 | $R_{2015}$ | 0.876 |
| $O_{2016}$ | 11353 | $C_2$ | 2.766 | $E_{2016}$ | 10030.896 | $R_{2016}$ | 1.132 |
| $O_{2017}$ | 7781 | $C_3$ | 3.718 | $E_{2017}$ | 9265.016 | $R_{2017}$ | 0.840 |
| $O_{2018}$ | 7680 | $C_4$ | 3.982 | $E_{2018}$ | 8072.972 | $R_{2018}$ | 0.952 |
| $O_{2019}$ | 5230 | $C_5$ | 3.959 | $E_{2019}$ | 4944.560 | $R_{2019}$ | 1.058 |
| $O_{2020}$ | 6046 | $C_6$ | 3.893 | $E_{2020}$ | 5653.621 | $R_{2020}$ | 1.070 |
| $O_{2021}$ | 5139 | $C_7$ | 3.909 | $E_{2021}$ | 4169.482 | $R_{2021}$ | 1.237 |
| $O_{2022}$ | 2470 | $C_8$ | 3.933 | $E_{2022}$ | 2271.128 | $R_{2022}$ | 1.089 |
| $O_{2023}$ | 917 | $C_9$ | 4.512 | $E_{2023}$ | 902.122 | $R_{2023}$ | 1.019 |
| $O_{2024}$ | 302 | $C_{10}$ | 3.791 | $E_{2024}$ | 232.195 | $R_{2024}$ | 1.308 |
| sum | 57029 | | | | 57029 | | |

From Table 13, we derive that $I_1=1$，$I_2$(SCIM without the union of Brazil and the Netherlands) = 1.058, which is almost the same as $I_2$(SCIM without China).

Now we come to the comparison of Brazil with the rest of the world (where also the Netherlands has been removed), and similarly for the Netherlands. For this, we compare the observed values for Brazil, respectively the Netherlands, with the expected values derived from the $C_k$-values of the world (in SCIM). Data are shown in Tables 14 and 15.

Table 14. Brazil

| $P_i$ Brazil | | $O_i$ Brazil | | $C_k$ World | | $E_i$ World | | $R_i$ external | |
|---|---|---|---|---|---|---|---|---|---|
| $P_{2015}$ | 14 | $O_{2015}$ | 388 | $C_1$ | 0.771 | $E_{2015}$ | 493.135 | $R_{2015}$ | 0.787 |
| $P_{2016}$ | 17 | $O_{2016}$ | 289 | $C_2$ | 2.766 | $E_{2016}$ | 534.353 | $R_{2016}$ | 0.541 |
| $P_{2017}$ | 18 | $O_{2017}$ | 559 | $C_3$ | 3.718 | $E_{2017}$ | 484.576 | $R_{2017}$ | 1.154 |
| $P_{2018}$ | 15 | $O_{2018}$ | 410 | $C_4$ | 3.982 | $E_{2018}$ | 344.814 | $R_{2018}$ | 1.189 |
| $P_{2019}$ | 11 | $O_{2019}$ | 150 | $C_5$ | 3.959 | $E_{2019}$ | 209.865 | $R_{2019}$ | 0.715 |
| $P_{2020}$ | 17 | $O_{2020}$ | 178 | $C_6$ | 3.893 | $E_{2020}$ | 258.155 | $R_{2020}$ | 0.690 |
| $P_{2021}$ | 19 | $O_{2021}$ | 184 | $C_7$ | 3.909 | $E_{2021}$ | 213.298 | $R_{2021}$ | 0.863 |
| $P_{2022}$ | 14 | $O_{2022}$ | 96 | $C_8$ | 3.933 | $E_{2022}$ | 101.406 | $R_{2022}$ | 0.947 |
| $P_{2023}$ | 11 | $O_{2023}$ | 37 | $C_9$ | 4.512 | $E_{2023}$ | 38.801 | $R_{2023}$ | 0.954 |
| $P_{2024}$ | 5 | $O_{2024}$ | 4 | $C_{10}$ | 3.791 | $E_{2024}$ | 3.835 | $R_{2024}$ | 1.043 |
| sum | | | 2295 | | | | 2682.238 | | |

Similarly, we find for the Netherlands (Table 15).

Table 15. The Netherlands

| $P_i$ the Netherlands | | $O_i$ the Netherlands | | $C_k$ World | | $E_i$ World | | $R_i$ external | |
|---|---|---|---|---|---|---|---|---|---|
| $P_{2015}$ | 9 | $O_{2015}$ | 246 | $C_1$ | 0.771 | $E_{2015}$ | 317.015 | $R_{2015}$ | 0.776 |
| $P_{2016}$ | 13 | $O_{2016}$ | 690 | $C_2$ | 2.766 | $E_{2016}$ | 408.623 | $R_{2016}$ | 1.689 |
| $P_{2017}$ | 10 | $P_{2017}$ | 1699 | $C_3$ | 3.718 | $E_{2017}$ | 269.209 | $R_{2017}$ | 6.311 |
| $P_{2018}$ | 7 | $P_{2018}$ | 233 | $C_4$ | 3.982 | $E_{2018}$ | 160.913 | $R_{2018}$ | 1.448 |
| $P_{2019}$ | 5 | $P_{2019}$ | 155 | $C_5$ | 3.959 | $E_{2019}$ | 95.393 | $R_{2019}$ | 1.625 |
| $P_{2020}$ | 5 | $P_{2020}$ | 121 | $C_6$ | 3.893 | $E_{2020}$ | 75.928 | $R_{2020}$ | 1.594 |
| $P_{2021}$ | 6 | $P_{2021}$ | 98 | $C_7$ | 3.909 | $E_{2021}$ | 67.357 | $R_{2021}$ | 1.455 |
| $P_{2022}$ | 1 | $P_{2022}$ | 2 | $C_8$ | 3.933 | $E_{2022}$ | 7.243 | $R_{2022}$ | 0.276 |
| $P_{2023}$ | 5 | $P_{2023}$ | 37 | $C_9$ | 4.512 | $E_{2023}$ | 17.637 | $R_{2023}$ | 2.098 |
| $P_{2024}$ | 9 | $P_{2024}$ | 9 | $C_{10}$ | 3.791 | $E_{2024}$ | 6.903 | $R_{2024}$ | 1.304 |
| sum | | | 3290 | | | | 1426.221 | | |

We see that, $I_1$(Brazil vs. World) = 0.856, $I_2$(Brazil vs. World) = 0.888. For the Netherlands, we have $I_1$(Netherland vs. World) = 2.307, $I_2$ (Netherlands vs. World) = 1.858. A yearly comparison of the external R-sequences of Brazil and the Netherlands is shown in Fig. 2.

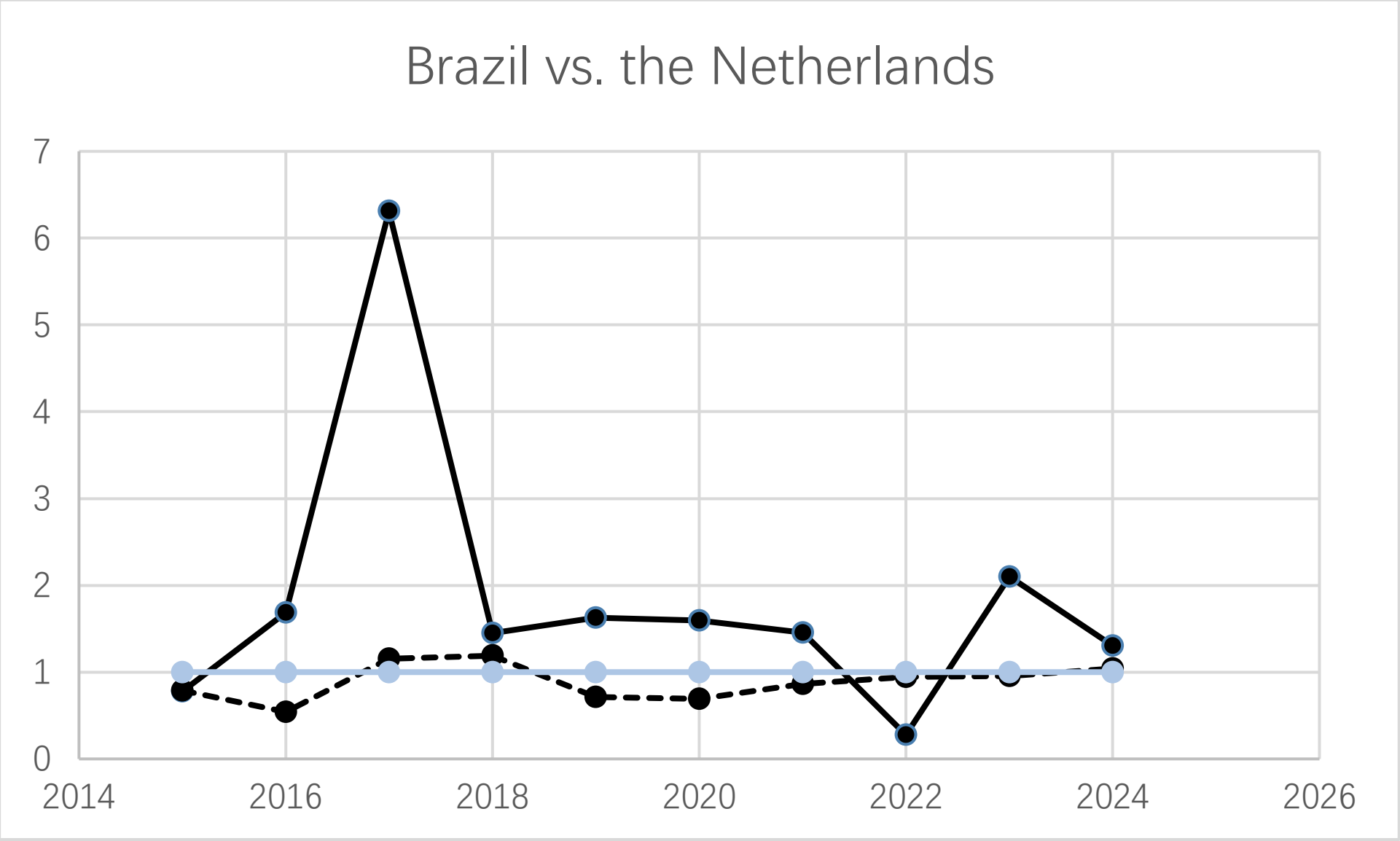

Fig.2. External rhythm sequences of Brazil (dotted line) and the Netherlands (full line) in SCIM

***Some comments on these examples***

We have illustrated that the indicators $I_1$ and $I_2$ for the external R-sequences can be compared among these three countries. The citation performance of the Netherlands is the best, followed by China and Brazil.

Observing Fig. 2, we observe that an extremely highly cited paper can influence the internal as well as the external R-sequence. The same holds for the I-indicators. The peak in the year 2017 is due to (van Eck & Waltman, 2017), which has received a total of 1519 citations (October 2025). This extremely highly cited paper makes that $O_{2017}$ accounts for over half of the total citations of the whole observed period. Therefore, $R_{2017}$ reached a high of 2.406, which directly dragged down the $R_i$ values of the other years of the Netherlands' internal R-sequence. Consequently, $I_2$ for the internal R-sequence of the Netherlands is as small as 0.873. This value is significantly lower than that of the other countries, as well as that of the world. This extremely highly cited paper also leads to $R_{2017}$ = 6.311 in the external R sequence, which is one of the main reasons why the indicators $I_1$=2.307 and $I_2$=1.858 for the Netherlands' external R-sequence are very high.

## Discussion

The practical example we have shown is, in a sense, a toy example (just one journal). Yet, it shows the potential of the external rhythm approach. Its true strength lies in the capacity to compare a large unit, like a major lab, with the broader discipline it is part of. In this, it uses all available data or better, all data that one wants to take into account. One may also start from a relatively small p-c table and shift this table over time, leading to a time series of R-sequences.

In our description of the data we have, for simplicity, used citations, counted as integers. Yet, we could as well count citations fractionally. Moreover, instead of citations associated with publications, the whole framework applies to any type of scores associated with publications, such as altmetric scores, on the condition that these scores are additive. Other variants of R-sequences (Liang, 2007) and other applications are possible. If the length of the citation window n increases, then all $C_k$ change.

The retracted article (Macháček & Srholec, 2021) has not been included. Retracted the next year (2022), it received, in this order, 15, 20, 10, 3, and 2 citations (period: 2021-2025).

We are aware of the distinction between the ratio of averages and the average of ratios (Opthof & Leydesdorff, 2010; Larivière & Gingras, 2011), yet we consider both $I_1$ and $I_2$ to be valuable in our context. Furthermore, because citation counts vary with publication year, with older publications generally accumulating more citations, one could even propose a weighted version of $I_2$. For now, however, we leave this line of inquiry to future work.

## Conclusion

The external indicator sequence and its average of ratios summary indicator introduced here are unique as they take all publication and citation information over a given period into account. Moreover, no overlapping of data occurs. Concretely, we provide a solution for the problem resulting from the occurrence of citation windows of different lengths. Having defined, constructed, and examined the external rhythm of an actor, we propose that this rhythm can serve as a potential citation indicator in science studies, applicable even to papers published in different years. For example, still using the example of countries, this indicator could be employed to compare the citation performance of individual countries with that of a collective group of countries (A), where each individual country represents an actor $B_q$, $A = \{B_q\}$, $q=1, 2, \ldots, m$. Similarly, one can also apply this indicator to study different actors, research groups, scientists, or journals.

**Limitations**

An important remark is that we assume that comparisons between an actor and the whole set, or between two actors within a larger set, are meaningful. For instance, if one actor $B_q$ is very dominant, so that $A \setminus B_q$ is very small, such a comparison is meaningless. A similar remark holds for a very small $B_q$. The same observation also holds for the union of $B_u$ and $B_v$ versus $A_{u,v}$.

**Declarations**

Competing interest. Ronald Rousseau is a member of the Distinguished Reviewer Board of Scientometrics.